\documentclass[intlimits,twoside,a4paper]{article}
\usepackage[eqsecnum]{cmpj3}

\usepackage[cp1251]{inputenc}
\usepackage{graphicx}
\usepackage{natbib}
\usepackage{bm}% bold math
\usepackage{epsfig}
\usepackage{xcolor}
\issue{2024}{27}{1}{13301}
\doinumber{10.5488/CMP.27.13301}

\title[Universal properties of branched copolymers in dilute solutions]%
{Universal properties of branched copolymers in dilute solutions}%

\author[K. Haydukivska, V. Blavatska]{K. Haydukivska\orcid{0000-0002-3118-7010}\refaddr{label1,label2}\thanks{Corresponding author: \email{wja4eslawa@icmp.lviv.ua}.},   
 V. Blavatska\orcid{0000-0001-6158-1636}\refaddr{label1,label3}
}
\addresses{
\addr{label1} Institute for Condensed Matter Physics of the National Academy of Sciences of Ukraine, 1 Svientsitskii Str., 79011 Lviv, Ukraine
\addr{label2} Institute of Physics, University of Silesia, 75 Pu\l{}ku Piechoty 1, 41-500 Chorz{\'o}w, Poland
\addr{label3}{Dioscuri Centre for Physics and Chemistry of Bacteria,
Institute of Physical Chemistry, Polish Academy of Sciences, 01-224 Warsaw, Poland}
}

\date{Received July 26, 2023, in final form October 18, 2023}

\begin{document}
\maketitle

\begin{abstract}
We analyze the universal conformational properties of complex copolymer macromolecules, based on two topologies: the \textit{rosette} structure containing $f_c$ linear branches and $f_r$ closed loops grafted to the central core, and 
the symmetric \textit{pom-pom} structure, consisting of a backbone linear chain terminated by two branching points with functionalities $f$. We assume that the constituent strands (branches) of these structures can be of two different  chemical species $a$ and $b$. Depending on the solvent conditions, the inter- or intrachain interactions of some links may vanish, which corresponds to $\Theta$-state of the corresponding polymer species.  Applying both the analytical approach within the frames of direct polymer renormalization and numerical simulations based on the lattice model of polymer,   we evaluated the set of parameters characterizing the size properties of constituent parts of two complex topologies  and estimated quantitatively the impact of interactions between constituent parts on these size characteristics.


\end{abstract}

\section{Introduction}

\begin{figure}[t!]
	\begin{center}
		\includegraphics[width=120mm]{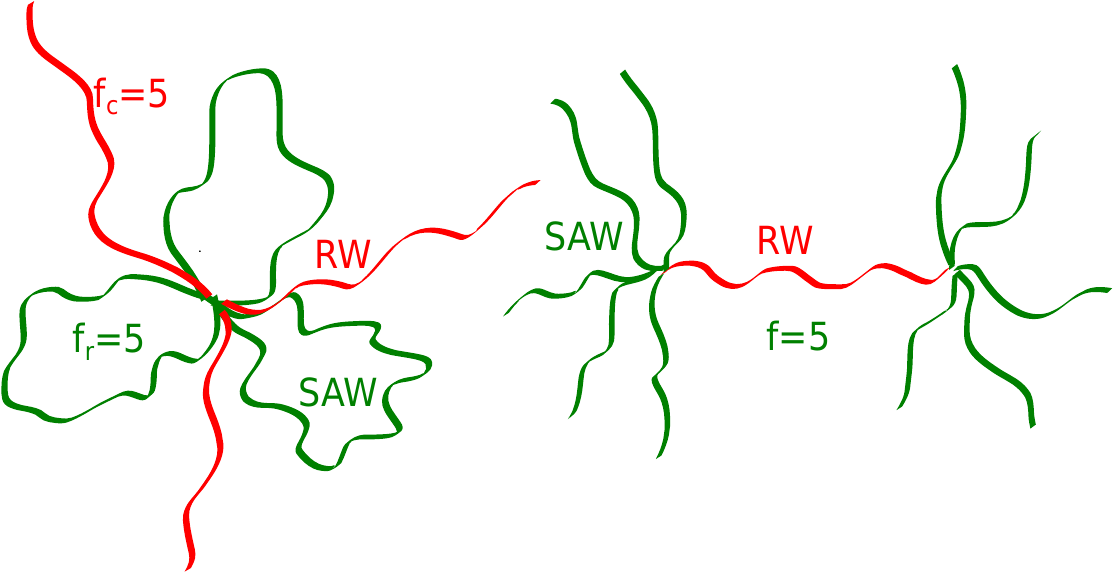}
		\caption{(Colour online) Schematic representation of rosette (left hand side) and symmetric pom-pom with $f_1=f_2=f$ (right hand side) copolymers. }
		\label{scheme} 
	\end{center}
\end{figure}

Polymers with  complex branching structure characterized by multiple main chain branching  
 attract a lot of attention in diverse bioapplications \cite{Gao04, Yates04, Voit09, Wang15, Cook20}. In particular, the complex topology of polymer macromolecules influences  the solution viscosity at a given concentration compared to a linear polymer of comparable molecular weight \cite{Schubert,Khabaz14,knauss2002,Fer2019}.
 The simplest representative of the class of branched polymers is the so-called star polymer with the single branching point having $f$ linear chains (branches) radiating from it  \cite{Zimm49}, {still attracting considerable attention of researchers \cite{Kalyuzhnyi19,Ilnytskyi08,Iln18}}.
A generalization of the star polymer, the so-called hybrid rosette structure is obtained when $f_r$ linear branches form closed loops, whereas $f_c$ branches remain linear \cite{Metzler15,Haydukivska18,Haydukivska20} (see figure~\ref{scheme} left-hand).  The polymer macromolecule with two branching points of functionalities $f_1$ anf $f_2$, which can be considered as two-star polymers with one common branch (the backbone) is known as pom-pom polymer \cite{Bishko,McLeish98,Haydukivska21,Haydukivska22a,Haydukivska23} (see figure~\ref{scheme} right-hand). { As more advanced polymer structures containing multiple banching points, we  should also mention  the  dendritic macromolecules resembling the structure of a tree with multiple repeating units, which have a wide range of potential applications \cite{Dend,Ilnytskyi08,Iln2}. }

Copolymers are complex molecules formed by linking the polymer chains of two different species. The synthesis of macromolecules with high molecular weight  enables the formation of new functional biomaterials with the potential for application in regenerative medicine, immunoengineering, imaging, and controlled drug delivery \cite{singh2015,schacher2017,ossorenen2012,meng2009,Mann21,Lin22}. Block copolymers are a specific class of polymer macromolecules containing 
subsequent blocks of two chemically distinct monomers $a$ and $b$ \cite{Hadjichristidis}. An essential feature of block-copolymer melts is their self-assembling
into lamellae and micelles and into more complex structures \cite{Matsen96,Mai12},
which provides potential applications in the fields such as producing dense and nanoporous membranes for gas separation and ultrafiltration \cite{Jackson10}, development of novel plastic materials~\cite{Bates} etc.

We concentrate on the conformational properties of complex polymer macromolecules in the regime of a weak solution. In particular, we consider a set of universal characteristics of  size and shape, which do not depend on the details of the chemical structures of a macromolecule. The most prominent example of such an observable is the size measure of the polymer macromolecule, the experimentally measurable gyration radius $R_g$ \cite{Burchard}, which scales with a molecular weight (number of monomers $N$) according to~\cite{deGennes,desCloiseaux}:
\begin{equation}
\langle R_g^2 \rangle = A N^{2\nu},
\label{eq_1}
\end{equation}
where $\nu$ is a scaling exponent that depends only on the space dimension and on the type of solvent in which the molecule is dissolved,   and $A$ is the amplitude that also contains the data about the topology (type of the branching) \cite{Zimm49}. Here and below, $\langle \ldots \rangle$ denotes averaging over an ensemble of all possible polymer conformations. For a chain in the regime of good solvent, where the repulsive excluded volume interactions between monomers play the main role, the scaling properties of linear polymer chain are perfectly captured by a model of self-avoiding random walk (SAW). In particular,  in $d=3$, one has $\nu_{\rm SAW}=0.58759700(40)$~\cite{Clisby16}. The model of random walk (RW) is exploited to describe the behaviour of a polymer chain in the regime of $\Theta$-solvent, when the excluded volume interactions between monomers can be neglected (Gaussian polymers). In this regime,   one has $\nu_{\rm RW}=1/2$ except lagarithmic corrections~\cite{Duplantier87}.

In $ab$ diblock copolymers, depending on the solvent and temperature regime, the polymer structure of some block (say $a$) can be in $\Theta$ state and may be effectively described as RWs (Gaussian chains), whereas the structure of $b$ type remains in good solvent regime modelled by  SAW.  Thus, the gyration radii of  blocks $a$ and $b$  scale according to~\eqref{eq_1} with exponents $\nu_{\rm SAW}$ and $\nu_{\rm RW}$, correspondingly \cite{Joanny84,Douglas87}, and there are three characteristic length scales in this case, governed by two types of inter-chain interactions $a$ and $b$ and intra-chain interaction $ab$.

 A number of studies were dedicated to the analysis of the universal conformation properties of diblock polymers \cite{vlahos1994,olaj1998,molina1995,Mcmullen1989,Binder2000}; the values of scaling exponents have been obtained both analytically \cite{vlahos1994,vlahos1995} and numerically \cite{vlahos1994,olaj1998}. In particular, it was observed  that the radii of gyration of each of the blocks are governed by the critical exponents that are defined by the type of their interaction with the solvent and are not dependent on the interaction with the other segments.  
The studies of star-copolymers \cite{rubio2000,Zifferer2010}  reveal the dependence of effective scaling exponents  on the number and type of branches. Thus, they are no longer topology independent. 

In the present study, we aim at analyzing the peculiarities of scaling behaviour of two examples of complex branched copolymer structures, based on rosette and pom-pom topologies (see figure~\ref{scheme}). We consider the constituent strands (branches) of these structures to be of two various species $a$ and $b$, so that one of the species can be in $\Theta$ state under a particular solvent and temperature condition, whereas the other one is in a good solvent regime.  The layout of the rest of the paper is as follows. We start with an analytical approach within  the frames of continuous chain model and use a direct renormalization approach in section~\ref{Theory}. The description of numerical algorithm along with general discussions of quantitative results obtained are given in section \ref{num}. We end up by giving conclusions and outlook in section \ref{con}.

\section{Theoretical approach}\label{Theory}

\subsection{Continuous chain model} \label{M}

Within the frames of continuous chain model approach, every  arm of a branched polymer is presented as a continious trajectory of length $L$  parameterized by radius vector $\vec{r_i}(s)$, where $s$ changes from $0$ to~$L$~\cite{Edwards}.
In the present  paper we consider pom-pom and rosette structures (see figure~\ref{scheme}).
The Hamiltonian model can be presented as:
\begin{eqnarray}
H&=& \frac{1}{2}\sum_{i=0}^{F}\,\int_0^{L} \rd s\,\left[\frac{\rd\vec{r_i}(s)}{\rd s}\right]^2+\frac{u_a}{2}\sum_{i=1}^{F_a}\int_0^{L}\rd s'\int_0^{L} \rd s''\,\delta(\vec{r_i}(s')-\vec{r_i}(s''))\nonumber\\
&+&\frac{u_b}{2}\sum_{i=1}^{F_b}\int_0^{L}\rd s'\int_0^{L} \rd s''\,\delta(\vec{r_i}(s')-\vec{r_i}(s''))+\frac{w_a}{2}\sum_{i=1}^{F_a}\sum_{i\neq j=1}^{F_a}\int_0^{L}\rd s'\int_0^{L} \rd s''\,\delta(\vec{r_i}(s')-\vec{r_j}(s''))\nonumber\\
&+&\frac{w_b}{2}\sum_{i=1}^{F_b}\sum_{i\neq j=1}^{F_b}\int_0^{L}\rd s'\int_0^{L} \rd s''\,\delta(\vec{r_i}(s')-\vec{r_j}(s''))\nonumber\\
&+&\frac{w_{ab}}{2}\sum_{i=1}^{F_b}\sum_{j=1}^{F_a}\int_0^{L}\rd s'\int_0^{L} \rd s''\,\delta(\vec{r_i}(s')-\vec{r_j}(s'')).
\label{H}
\end{eqnarray}
Here, $F$ is a total number of constituent chains and $F_a,\, F_b$ are the number of chains in subgroups $a$ and~$b$. $u_a$ and $u_b$ are the coupling constants for the excluded volume interaction between the points on the same chain of ether type $a$ or $b$, $w_a$ and $w_b$ are the coupling constants for the excluded volume interaction between the points on different trajectories of the same type (either $a$ type with $a$ type or $b$ type with $b$ type) and $w_{ab}$ is a coupling constant for the excluded volume interaction between chains of different types. With $u_a=u_b=w_a=w_b=w_{ab}$ we restore a case of homopolymer. 

The topologies of the macromolecules are accounted for in the definition of partition functions. For pom-pom polymer with $F=f_1+f_2+1$ it reads:
\begin{eqnarray}
&&Z_{f_1,f_2}=\frac{1}{Z_0^{{\rm pom-pom}}}\int\,D\vec{r}(s)\prod_{i=1}^{f_1}\prod_{j=1}^{f_2}\,\delta(\vec{r_i}(0)-\vec{r_0}(0))\,
\delta(\vec{r_j}(0)-\vec{r_0}(L))\,{\rm e}^{-H}.
\label{ZZ}
\end{eqnarray}
Here, the backbone chain is considered as $0$th [parametrized by $r_0(L)$], the set of $\delta$-functions describes the fact that $f_1+1$ trajectories start at one end point of the backbone and $f_2$ trajectories start at its other end point $\vec{r_0}(L)$. We consider the case when $f_1=f_2=f$,  all arms are of the same type $a$ ($F_a=2f$) and the backbone is of type $b$ ($F_b=1$). 

For the case of rosette polymer, the partition function reads:
\begin{equation}
Z_{\rm rosette}=\frac{1}{Z_0}\int D\vec{r} \prod_{i=1}^{f_c+f_r}\delta(\vec{r_i}(0))\prod_{j=1}^{f_r} \delta(\vec{r_j}(L)-\vec{r_j}(0))\, {\rm e}^{-H_0}.
\end{equation}
Here, $f_c$ and $f_r$ being the number of open and closed trajectories, correspondingly, with chains being of type $a$ and rings of type $b$. 

\subsection{Direct renormalization method} \label{Met}

To describe the universal properties of a continuous chain model one needs to perform  renorma\-li\-zation. In this work we use a direct renormalization scheme developed by des Cloiseaux \cite{desCloiseaux}.

The aim of this method is to eliminate the divergences that appear in the limit of infinitely long trajectories by introducing a  set of renormalization factors. All of the universal parameters are finite at the so-called fixed points (FP). The FPs of the renormalization are defined as common zeros for the $\beta$-functions:
\begin{eqnarray}
\beta_{u_{i,R}} &=&\epsilon u_{i,R}-8u^2_{i,R}=0, \label{b1}\\
\beta_{w_{i,j,R}} &=&\epsilon w_{i,j,R}-\frac{(L_i+L_j)^2}{L_iL_j}w^2_{i,j,R}\nonumber\\ &-&2w_{i,j,R}(u_{i,R}+u_{j,R})=0. 
\label{b2}
\end{eqnarray}
Here, $u_{i,R}$ is renormalized excluded volume interaction constant of the $i$-th chain and $w_{i,j,R}$ is a constant between chains $i$ and $j$~\cite{Haydukivska19}.

Fixed points of the model for a case of all $L_i=L$ are:
\begin{eqnarray}
&&\widetilde{u}_{i,R} = 0, \quad \widetilde{u}_{i,R} = \frac{\epsilon}{8},
\end{eqnarray}
and for $\widetilde{w}_{i,j,R}$:
\begin{eqnarray}
&&\widetilde{w}_{i,j,R} = 0,  \quad \forall \,\widetilde{u}_i, \, \widetilde{u}_j, \label{1}\\
&&\widetilde{w}_{i,j,R} = \frac{\epsilon}{4}, \quad \widetilde{u}_i=\widetilde{u}_j=0,\label{2} \\
&&\widetilde{w}_{i,j,R} = \frac{3\epsilon}{16}, \quad \widetilde{u}_i\neq \widetilde{u}_j, \label{3}\\
&&\widetilde{w}_{i,j,R} = \frac{\epsilon }{8},\quad \widetilde{u}_i=\widetilde{u}_j= \frac{\epsilon}{8}.\label{4}
\end{eqnarray}

\subsection{Results}
\paragraph{Partition function of the pom-pom structure.}

\begin{figure}[b!]
	\begin{center}
		\includegraphics[width=73mm]{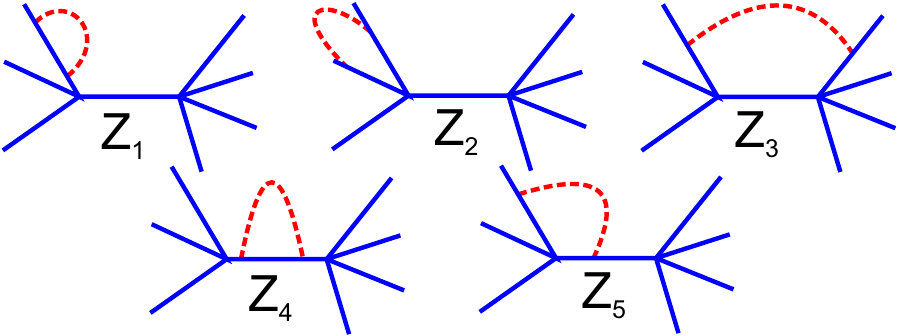}
		\caption{(Colour online) Diagrammatic representations of contributions into the  partition function of the pom-pom structure up to the first order of perturbation theory expansion in coupling constants.
			The solid lines are schematic presentations of polymer strands, and the dash line represents a two monomer excluded volume interaction.}
	\label{fig:2}
	\end{center}
\end{figure}

We start our calculations from considering the partition function of copolymer pom-pom structure. These calculations are made within the perturbation theory framework using diagramatic technique \cite{desCloiseaux} to calculate different contributions into the  partition function. The corresponding diagrams are presented in figure \ref{fig:2}. We consider only linear terms of the expansion in the coupling constants $u_i$, $w_i$. The contributions from the diagrams can be presented as:
\begin{eqnarray}
&&Z_1=\frac{u_a(2\piup)^{-d/2}L^{2-({d/2})}}{\left(1-\frac{d}{2}\right)\left(2-\frac{d}{2}\right)},\\
&&Z_2=\frac{w_a(2\piup)^{-d/2}L^{2-({d/2})}\left[2^{2-({d}/{2})}-2\right]}{\left(1-\frac{d}{2}\right)\left(2-\frac{d}{2}\right)},\\
&&Z_3=\frac{w_a(2\piup)^{-d/2}L^{2-({d/2})}\left[3^{2-({d}/{2})}-2(2)^{2-({d}/{2})}+1\right]}{\left(1-\frac{d}{2}\right)\left(2-\frac{d}{2}\right)},\\
&&Z_4=\frac{u_b(2\piup)^{-d/2}L^{2-({d/2})}}{\left(1-\frac{d}{2}\right)\left(2-\frac{d}{2}\right)},\\
&&Z_5=\frac{w_{ab}(2\piup)^{-d/2}L^{2-({d/2})}\left[2^{2-({d}/{2})}-2\right]}{\left(1-\frac{d}{2}\right)\left(2-\frac{d}{2}\right)}.
\end{eqnarray}
 Each of the diagrams should be accounted for with a corresponding  pre-factor, which equals $2f$ for diagram~$Z_1$, $f(f-1)$ for $Z_2$ so that contributions from both side-stars are included. Diagram $Z_3$ has a pre-factor $f^2$. The diagrams $Z_4$ and $Z_5$ that account for the contributions from the interactions related to the backbone have pre-factors $1$ and $2f$, correspondingly.

Analytical expressions corresponding to the diagrams are presented as functions of space dimension~$d$, chain length $L$ and coupling constants. The results of expansion over the deviation from the upper critical dimension $\epsilon=4-d$ read:
\begin{eqnarray}
&&Z_1=\widetilde{u}_{a}\left(-\frac{2}{\epsilon}-1\right),\\
&&Z_2=\widetilde{w}_{a}\left(\frac{2}{\epsilon}+1-\ln2 \right),\\
&&Z_3=\widetilde{w}_{a}\left(2\ln2 -\ln3 \right),\\
&&Z_1=\widetilde{u}_{b}\left(-\frac{2}{\epsilon}-1\right),\\
&&Z_2=\widetilde{w}_{ab}\left(\frac{2}{\epsilon}+1-\ln2 \right).
\end{eqnarray}
Here, $\widetilde{u}_a=u_a(2\piup)^{-{d}/{2}}L^{2-({d}/{2})}$, $\widetilde{w}_a=w_x(2\piup)^{-{d}/{2}}L^{2-({d}/{2})}$ are dimensionless coupling constants.
The final expression for the partition function in one-loop approximation reads:
\begin{eqnarray}
Z_{f,f}&=&1-\left(2f\widetilde{u}_{a}+\widetilde{u}_{b}\right)\left(-\frac{2}{\epsilon}-1\right)-\left[f(f-1)\widetilde{w}_{a}+2f\widetilde{w}_{ab}\right]\left(\frac{2}{\epsilon}+1-\ln 2\right)\nonumber\\
&+&f^2\widetilde{w}_{a}\left(2\ln 2-\ln 3 \right). \label{Zfinal}
\end{eqnarray}
Even so, we are not interested in the scaling properties of partition function, although it still plays an important role in calculating other observables since their averaging in general is defined as:
\begin{eqnarray}
&&\langle (\ldots) \rangle = \frac{1}{{ Z_{f,f}}}\prod_{i=1}^{f}\prod_{j=1}^{f}\,\int\,D\vec{r}(s)\delta(\vec{r_i}(0)-\vec{r_0}(0))\delta(\vec{r_j}(0)-\vec{r_0}(L))\,{\rm e}^{-H}(\ldots).
\end{eqnarray}

\paragraph{Partition function of the rosette structure.}

\begin{figure}[t!]
	\begin{center}
		\includegraphics[width=73mm]{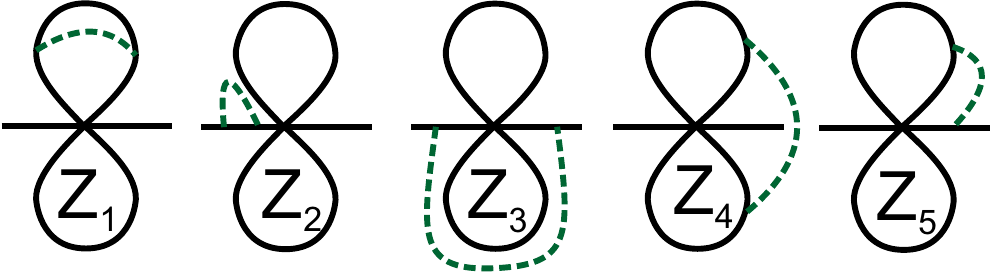}
		\caption{Diagrammatic representations of contributions into the partition function of rosette polymer structure up to the first order of perturbation theory in the  coupling constants. The solid lines are schematic presentations of polymer strands, and the dash line represents a two monomer excluded volume interaction.}
		 \label{fig:2a}
	\end{center}
\end{figure}

Similarly, we conduct the calculations for the rosette polymers (see the diagrammatic presentation  of figure~\ref{fig:2a}). The analytical expressions corresponding to the diagrams read:
\begin{eqnarray}
Z_1&=&-2\widetilde{u}_{b}\,(2\piup L)^{-\frac{d}{2}f_r}\frac {\Gamma \left( 2-\frac{d}{2} \right)^{2}}{\left(d-2\right) \Gamma  \left( 3-d \right) }, \nonumber \\
 Z_2&=&\widetilde{u}_{a}\,(2\piup L)^{-\frac{d}{2}f_r}\frac{4}{(4-d)(2-d)}, \nonumber \\
Z_3&=&\widetilde{w}_{a}\,(2\piup L)^{-\frac{d}{2}f_r}\frac{4(2^{2-d/2}-2)}{(4-d)(2-d)}, \nonumber \\
Z_4&=& \widetilde{w}_{b}\,(2\piup L)^{-\frac{d}{2}f_r}\left\{-\frac{1}{8}\,{\frac {{2}^{d}\sqrt {\piup }\Gamma  \left( 1-\frac{d}{2} \right) }{
\Gamma  \left(\frac{5-d}{2} \right) }} \right.\nonumber \\
&+&\left.\frac{1}{3}\,{\frac {{2}^{d-1}{5}^{-\frac{d}{2}}\left[
		{\mbox{$_2$F$_1$}\left(\frac{3}{2},\frac{d}{2};\,\frac{5}{2};\,\frac{1}{5}\right)}-3\,{\mbox{$_2$F$_1$}\left(\frac{1}{2},\frac{d}{2};\,\frac{3}{2};\,\frac{1}{5}\right)}\right]}{d-2}}\right\},\nonumber\\
Z_5&=&\widetilde{w}_{ab}\,(2\piup L)^{-\frac{d}{2}f_r} {\frac {{2}^{{(d-1)}/{2}}\sqrt {\piup }\,
{\mbox{$_2$F$_1$}\left(\frac{1}{2},\frac{d-1}{2};\,\frac{3}{2};\,\frac{1}{2}\right)}\Gamma  \left( 1-\frac{d}{2}
 \right) }{\Gamma  \left(\frac{3-d}{2} \right) }}.
\end{eqnarray}

Each of these diagrams is taken with an additional pre-factor: for the diagrams $Z_1$ and $Z_2$ it is correspondingly the number of rings $f_r$ and the number of chains $f_c$; diagrams $Z_3$ and $Z_5$ should be accounted for each pair of chains [$f_c(f_c-1)/2$] or rings [$f_r(f_r-1)/2$] and the diagram $Z_4$ for each pair of one chain and one ring $f_cf_r$. Performing the expansion of corresponding expressions in  $\epsilon=4-d$,  we obtain:
\begin{eqnarray}
Z^{f_c,f_r}&=&(2\piup L)^{-\frac{d}{2}f_r}\Bigg\{1-\frac{f_c^2\widetilde{w}_a+4f_cf_r\widetilde{w}_{ab}+2f_r^2\widetilde{w}_b-2f_c\widetilde{u}_a-f_c\widetilde{w}_a+4f_r\widetilde{u}_b-2f_r\widetilde{w}_b}{\epsilon}\nonumber\\
&+&f_c\widetilde{u}_a+2f_r\widetilde{u}_b+\frac{f_c(f_c-1)}{2}(\ln 2-1)\widetilde{w}_a+\widetilde{w}_bf_r(f_r-1)\nonumber\\
&\times&\left[\sqrt{2}\int_0^1\,\rd t\frac{\ln(2-t)}{\sqrt{t}(t-2)\sqrt{4-2t}}+1\right]+\frac{2}{5}f_cf_r\widetilde{w}_{ab}\left[\sqrt{5}\ln \left(\sqrt{5}+3\right)-\sqrt{5}\ln 2-5\right]\Bigg\}.
\end{eqnarray}
This expression is used in calculations of averaging of the observables considered below, with the averaging  defined as:
\begin{eqnarray}
&&\langle (\ldots) \rangle = \frac{1}{ Z}\,\int\,D\vec{r}(s)F[\delta]\,{\rm e}^{-H}(\ldots).
\end{eqnarray}
Here, $F[\delta]$ is a set of $\delta$-functions defining the topology.

\paragraph{Size characteristics.}
\begin{figure}[t!]
	\begin{center}
		\includegraphics[width=63mm]{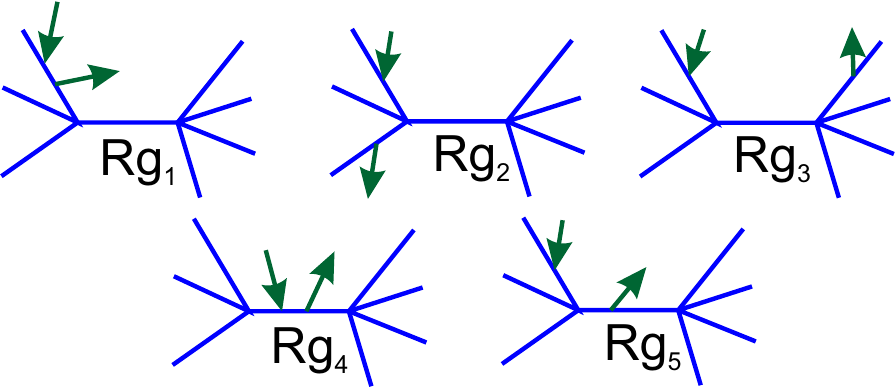}
		\includegraphics[width=63mm]{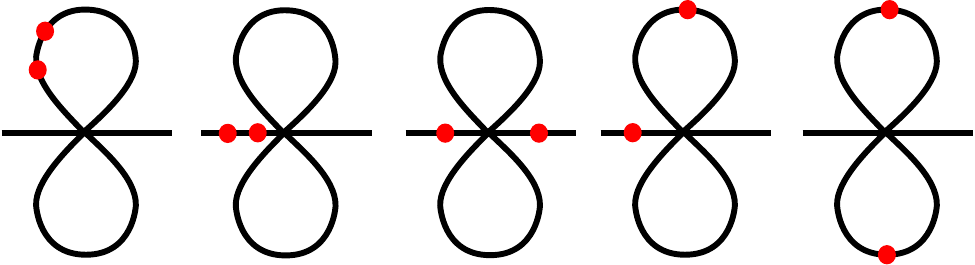}
		\caption{(Colour online) Diagrammatic representation of contributions into $\xi(\vec{k})$ in Gaussian approximation. The solid lines are schematic presentations of polymer strands, each of length $L$, and arrows represent the so-called restriction points $s_1$ and $s_2$.}
		 \label{fig:3}
	\end{center}
\end{figure}

Within the continuous chain model, the gyration radius is defined as:
\begin{eqnarray}
&&{\langle {R^2_{g}}\rangle} = \frac{1}{2L^2(F)^2}\sum_{i,j=0}^{F}\int_0^{L}\int_0^{L} \rd s_1\,\rd s_2 \langle(\vec{r}_i(s_2)-\vec{r}_j(s_1))^2\rangle. \label{totalRg}
\end{eqnarray}
In the case of copolymer structure, containing constituent strands of different chemical nature, the gyration radius (\ref{totalRg})  does not have an analytically defined scaling. Instead, each strand  has its own scaling behaviour as well as the contributions from correlations between chains lead to its own scaling behaviour \cite{Haydukivska19,olaj1998}. We have:
\begin{equation}
{\langle {R^2_{g}}\rangle} ={\langle {r^2_{g,a}}\rangle}+{\langle {r^2_{g,b}}\rangle}+{\langle {r^2_{g,ab}}\rangle}, \label{Rgsum}
\end{equation}
where the definition of components  may depend on the topology of polymer structure. In the case of rosette polymers, diagrams with restriction points on the linear branches give contribution into ${\langle {r^2_{g,a}}\rangle}$; the one with both restriction points on the closed branches (loops) give a contribution into ${\langle {r^2_{g,b}}\rangle}$ and the rest --- into ${\langle {r^2_{g,ab}}\rangle}$. Similar is the situation in the case of pom-pom structure with the exception of the diagrams with restriction points on different poms, which give the contribution into ${\langle {r^2_{g,ab}}\rangle}$. 
A full list of diagrams separated into the proper groups is provided in the Appendix.

To calculate the corresponding analytical expression, an identity is used 
\begin{eqnarray}
&&\langle(\vec{r}_i(s_2)-\vec{r}_j(s_1))^2\rangle = - 2 \frac{\rd}{\rd|\vec{k}|^2}\xi(\vec{k})_{\vec{k}=0},\nonumber\\
&&\xi(\vec{k})\equiv \langle\exp\{-\ri\vec{k}(\vec{r}_i(s_2)-\vec{r}_j(s_1))\}\rangle.
\end{eqnarray}
The final expression for gyration radius of backbone of a pom-pom structure reads:
\begin{eqnarray}
&&\langle r^2_{g,\,{\rm b}}\rangle_{pp}=\frac{dL}{6}\left[1+\frac{2\widetilde{u}_b}{\epsilon}+\left(\frac{35}{4}-12\ln2\right)f \widetilde{w}_{ab}+\frac{f^2 \widetilde{w}_a}{12}-\frac{13}{12}\widetilde{u}_b\right],\label{r_backbone}
\end{eqnarray}
gyration radius of each ``pom'' is given by:
 \begin{eqnarray}
{\langle{r^2_{g,a}}\rangle}_{pp}&=&\frac{dLf(3f-2)}{3}\Bigg(1+\frac{2\widetilde{u}_a}{\epsilon}+\frac{1}{24(3f-2)}\left\{\widetilde{w}_{ab}[144\ln2(2f-3)\right.\nonumber\\
&-&3(52f-87)]{+}\widetilde{w}_a\left[12\ln2(3f^3{-}78f^2{+}119f+40){-}
36f^3{+}864\ln3f(f-2)\right.\nonumber\\
&-&\left.\left. 318-232f^2+837f\right]-2\widetilde{u}_a(42f-29)\right\}\Bigg),\label{r_pom}
\end{eqnarray}
and the remaining ``mixed'' part of the gyration radius will read:
\begin{eqnarray}
&&{\langle {r^2_{g,ab}}\rangle}_{pp} = dL2f^2(f{+}1)\Bigg(1{+}\frac{\widetilde{u}_a{+}\widetilde{u}_b}{\epsilon}{+}\frac{1}{72(f+1)}\left.\left\{2\widetilde{w}_a[12\ln 2(3f^2-28f-6)\right.\right.\nonumber\\
&&+ \,9\ln 3 \, f(8f+25)+3f^3-79f^2-20f+39]-\widetilde{w}_{ab}[14f^2+76f+21\nonumber\\
&&-\left.72\ln3 \, f(f-1)+24\ln 2 \,(3f^2-15f-4)]-42\widetilde{u}_a(f+1)-6\widetilde{u}_b(6f+7)\right\}\Bigg)\label{r_mix}.
\end{eqnarray}
Similarly, we derive the expressions for the case of rosette polymer structure:
\begin{eqnarray}
\langle r^2_{g,\,{\rm a}}\rangle_{r}&=&\frac{f_r(3f_c-2)dL}{6}\Bigg\{1+\frac{2\widetilde{u}_a}{\epsilon}+\frac{(f_c-1) [16 \ln 2 (3 f_c-5)-26 f_c+53] \widetilde{w}_a}{4(3 f1-2)}\nonumber\\
&-&\frac{\widetilde{u}_a (42 f_c-29)}{12(3 f_c-2)}+\frac{f_r\widetilde{u}_{ab}}{25(3f_c-2)}\left[\sqrt{5}\arctan\big(5^{-\frac{1}{2}}\big)\left(60f_c^2+132f_c-199\right)\right.\nonumber\\
&+&\left.10\sqrt{5}\ln 2(3f_c^2-3f_c+1)-110f_c+150-10\sqrt{5}\ln(\sqrt{5}+3)(3f_c^2-3f_c+1)\right]\Bigg\},\\
\langle r^2_{g,\,{\rm b}}\rangle_{r}&=&\frac{f_r(2f_r-1)dL}{12}\Bigg(1+\frac{2\widetilde{u}_b}{\epsilon}+\frac{\widetilde{w}_b(f_r-1)}{8(2f_r-1)}\Big\{\sqrt{2}\arctan\big(2^{-\frac{1}{2}}\big)(34f_r-59)\nonumber\\
&+&\int_0^1 \frac{16\sqrt{2}\ln(2-t)(t^2-3t+1)}{(\sqrt(t)(t-2)\sqrt{4-2t})}\rd t-\int_0^1 \frac{768t^2{\rm arctan}\big[(4t^2-4t-1)^{-\frac{1}{2}}\big](t-1)^2(f_r-2)}{(4t^2-4t-1)^{\frac{5}{2}}}\rd t\nonumber\\
&-&\int_0^1 \frac{64(t-1){\rm arctan}\big[(-4t^2+4t+1)^{-\frac{1}{2}}\big]}{(-4t^2+4t+1)^{\frac{5}{2}}}(30t^4-63t^3+19t^2+14t+2)\rd t\Big\}\nonumber\\
&+&\frac{\widetilde{w}_{ab}f_c}{8400(2f_r-1)}\left\{2\sqrt{5}\arctan(5^{-\frac{1}{2}})(6720f_r^2+76763f_r\right.-77435)\nonumber\\
&-&3360\sqrt{5}\ln(\sqrt{5}+3)f_r(2f_r-1)\nonumber\\
&+&\left.155567+70\ln 2[48(2f_r^2-1)\sqrt{5}-197(f_r-1)]-157247f_r\right\}\Bigg),\\
\langle r^2_{g,\,{\rm ab}}\rangle_{r}&=&\frac{f_r f_c d L}{12}\Bigg\{1+\frac{\widetilde{u}_b+3\widetilde{u}_a}{2\epsilon}-\frac{7}{8}\widetilde{u}_b
+\frac{\widetilde{w}_b}{8}(f_c-1)(24\ln 2-13)\nonumber\\
&+&\frac{\widetilde{w}_{ab}}{100} \big[\sqrt{5}\arctan\big(5^{-\frac{1}{2}}\big)
(80f_cf_r+32f_c+192f_r+7)-40\sqrt{5}\ln(\sqrt{5}+3)f_cf_r\nonumber\\
&+&40\sqrt{5}\ln(2)f_cf_r-10f_c-110f_r+40\big]\nonumber\\
&-&\frac{(f_r-1)\widetilde{w}_a}{32}\Bigg[192\int_1^{\sqrt{2}}\frac{\arctan\big(s^{-1}\big)(s-1)^2(s+1)^2}{\sqrt{2-s^2}s^4}\rd s \nonumber\\
&+&29\sqrt{2}\arctan\big(2^{-\frac{1}{2}}\big)-84\ln 2+8\Bigg]\Bigg\}.
\end{eqnarray}

\paragraph{Scaling exponents}
\paragraph{Size characteristics exponents.}

As it was mentioned above, in the case of copolymer structure, we have three characteristic length scales, governed by two types of inter-chain interactions $a$ and $b$ and intra-chain interaction $ab$, so that  the  segments of different species as well as correlations between segments are governed by different scaling exponents. Within the continuous chain model, the size exponents may be calculated using the expression \cite{desCloiseaux}:
\begin{eqnarray}
2\nu_x-1&=&\frac{\epsilon}{2}\left(\widetilde{u}_a\frac{{\rm d} \ln \langle r_{g,x}^2\rangle}{{\rm d}\ln\widetilde{u}_a}+\widetilde{u}_b\frac{{\rm d} \ln \langle r_{g,x}^2\rangle}{{\rm d}\ln\widetilde{u}_b}+\widetilde{w}_a\frac{{\rm d} \ln \langle r_{g,x}^2\rangle}{{\rm d}\ln\widetilde{w}_a}\right.\nonumber\\
&+&\left.\widetilde{w}_b\frac{{\rm d} \ln \langle r_{g,x}^2\rangle}{{\rm d}\ln\widetilde{w}_b}+\widetilde{w}_{ab}\frac{{\rm d} \ln \langle r_{g,x}^2\rangle}{{\rm d}\ln\widetilde{w}_{ab}}\right).\label{nu}
\end{eqnarray}
 The estimates for the critical exponents governing correspondingly each of the three terms in (\ref{Rgsum}) for the case of pom-pom structure read:
\begin{eqnarray}
&&\nu^{pp}_{b}=\frac{1}{2}\left(1+\widetilde{u}_b\right),\label{nu_b}\\
&&\nu^{pp}_{a}=\frac{1}{2}\left(1+\widetilde{u}_a\right),\label{nu_a}\\
&&\nu^{pp}_{ab}=\frac{1}{2}\left(1+\frac{\widetilde{u}_b+\widetilde{u}_a}{2}\right),\label{nu_m}
\end{eqnarray}
whereas for the case of rosette polymer we get:
\begin{eqnarray}
&&\nu^{r}_{b}=\frac{1}{2}\left(1+\widetilde{u}_b\right),\label{nu_br}\\
&&\nu^{r}_{a}=\frac{1}{2}\left(1+\widetilde{u}_a\right),\label{nu_ar}\\
&&\nu^{r}_{ab}=\frac{1}{2}\left(1+\frac{\widetilde{u}_b+3\widetilde{u}_a}{4}\right).\label{nu_mr}
\end{eqnarray}
Similarly, like it was observed in the case of block copolymers in \cite{Haydukivska19, olaj1998, vlahos92}, scaling exponents $\nu^{x}_{a}$ and $\nu^{x}_{b}$ remain unchanged by the  presence of the second species in the polymer structure and do not depend on the  interaction between different species $\widetilde{w}_x$. 
 With $\widetilde{u}_b=\widetilde{u}_a$, one recovers the homopolymer behaviour.

Though the universal exponent of a total gyration radius  of the whole copolymer structure cannot be defined, we can derive an expression for the gyration radius ${\langle {R^2_{g}}\rangle}$ by simply adding all the diagrams together. In general, it will read:
\begin{equation}
{\langle {R^2_{g}}\rangle} = {\langle {R^2_{g}}\rangle}_0\left[1+(\ldots)\right],
\end{equation}
with ${\langle {R^2_{g}}\rangle}_0$ being the Gaussian approximation and $\left[1+(\ldots)\right]$ a swelling factor in one loop approximation in the coupling constants. According to the definition (\ref{nu}) we may evaluate an effective critical exponent for the cases of pom-pom and rosette structures, correspondingly:
\begin{eqnarray}
\nu^{pp}_{eff}&=&\frac{1}{2}\left(1+\frac{12f^2\widetilde{u}_a+6f^2\widetilde{u}_b+2f\widetilde{u}_a+6f\widetilde{u}_b+\widetilde{u}_b}{18f^2+8f+1}\right),\label{nup}\\
\nu^{r}_{eff}&=&\frac{1}{2}+\frac{f_c(3f_c+3f_r-2)\widetilde{u}_a}{6f_c^2+8f_cf_r+2f_r^2-4f_c-f_r}\nonumber\\
&+&\frac{f_r(2f_c+2f_r-1)\widetilde{u}_b}{2(6f_c^2+8f_cf_r+2f_r^2-4f_c-f_r)}.\label{nur}
\end{eqnarray}
Again, for $u_a=u_b$, we recover an expression for an exponent of a homopolymer. These exponents are not properly defined. They may have a more practical application, since they can be related to the one observed in the experiment. Such calculations were previously performed in 
simulations in reference~\cite{rubio2000} for star copolymers for which the topology dependent effective exponents were obtained. 

\subsection{Size characteristics}

Though the interactions between different strands do not influence the scaling exponents, they do have an impact on the size characteristics which can be shown through consideration of the size ratios. In the present work we consider size ratios $p^{pp}_g$, $ p^{r}_g$ defined as
\begin{eqnarray}
    &&p^{pp}_g=\frac{\langle r^2_{g,\,{\rm b}}\rangle_{pp}}{\langle R^2_{g,\,{\rm chain}}\rangle}, \label{ratiob} \\
    &&p^{r}_g=\frac{\langle r^2_{g,\,{\rm arm}}\rangle_{r}}{\langle R^2_{g,\,{\rm chain}}\rangle},\label{ratior}  
\end{eqnarray}
where $\langle r^2_{g,\,{\rm b}}\rangle_{pp}$  and $\langle r^2_{g,\,{\rm arm}}\rangle_{r}$  are correspondingly  radii of gyration of the backbone of pom-pom polymer and a linear branch in rosette polymer, and  
${\langle R^2_{g,\,{\rm chain}}\rangle}$ is the radius of gyration of individual linear chain of  the same molecular mass. 

\begin{figure}[t!]
	\begin{center}
		\includegraphics[width=74mm]{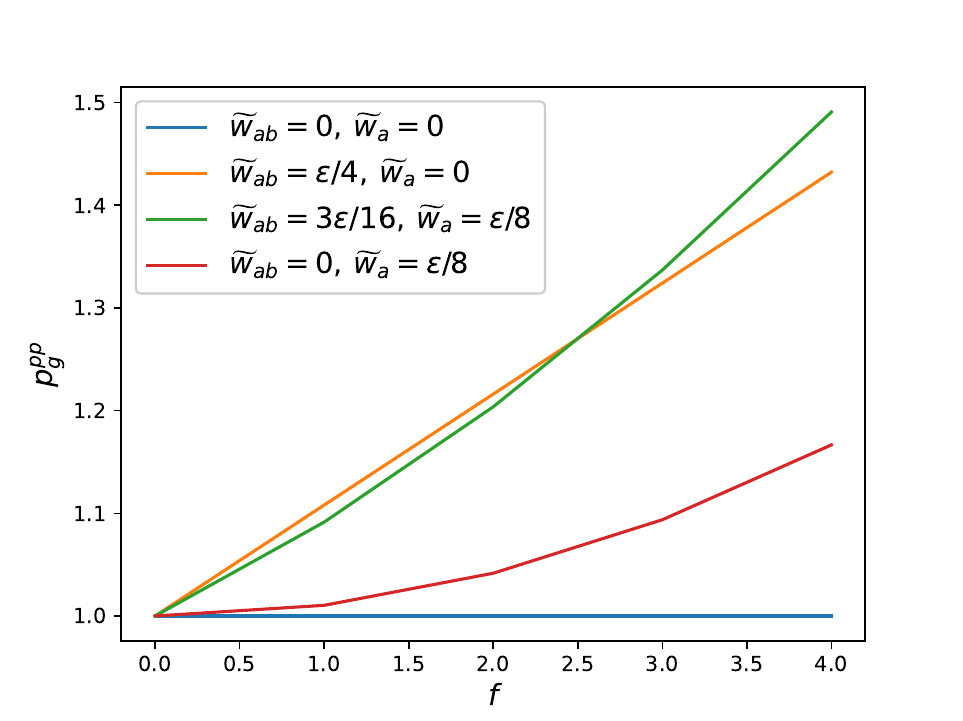}
            \includegraphics[width=74mm]{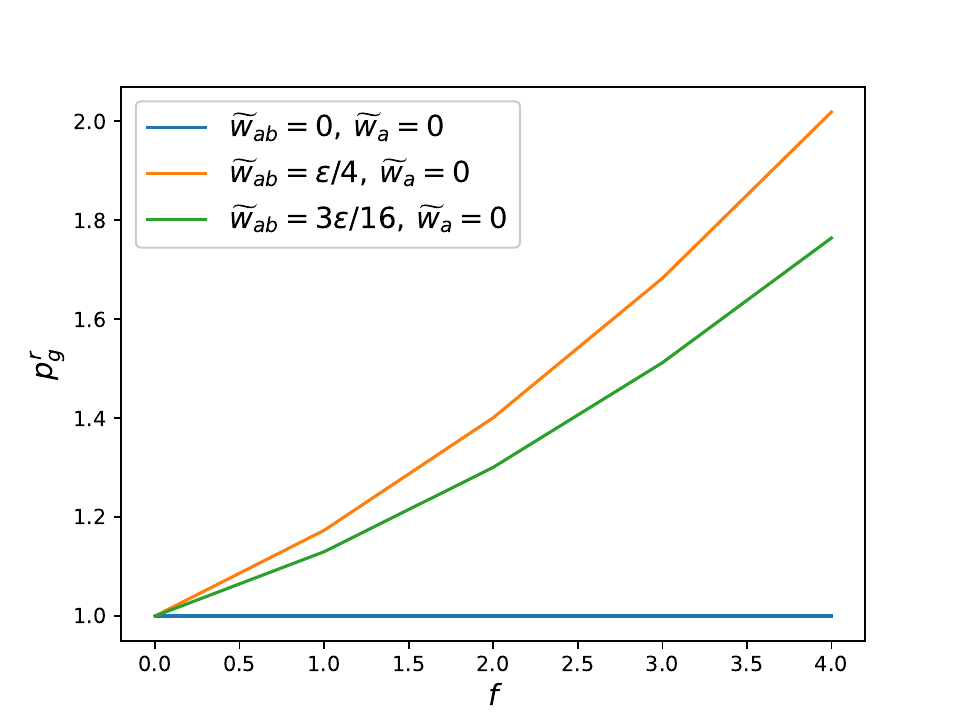}
		\caption{(Colour online) On the left: Size ratio $p^{pp}_g$ as a function of the  branching parameter $f$ for a RW backbone and ether RW or SAW poms. On the right: Size ratio $p^{r}_g$ for an linear RW arm in rosette as a function of $f_c=f_r=f$ and rings being either RW or SAW.}
		\label{pg} 
	\end{center}
\end{figure}

Substituting expression (\ref{r_backbone}) into the numerator of (\ref{ratiob}) and the same expression with $f=0$ into the denominator and performing a series expansion of the denominator in respect to the coupling constants, we get the expression:
\begin{equation}
    p^{pp}_g=1+\left(\frac{35}{4}-12\ln2 \right)f \widetilde{w}_{ab}+\frac{f^2 \widetilde{w}_a}{12}. \label{sizepp}
\end{equation}
Let us consider the case,  when the backbone of pom-pom polymer is of a RW type ($\widetilde{u}_b=0$), and the size ratio with, correspondingly, the RW linear chain is considered. The results of evaluation of the size ratio (\ref{sizepp}) at different possible interactions within the side poms and between side poms and backbone are presented in figure~\ref{pg} (left-hand) as functions of the branching parameter $f$. The case when side poms are also RWs and there is not mutual interaction between backbone and side poms ($\widetilde{w}_{ab}=\widetilde{w}_a=0$) is trivial and gives  $p^{pp}_g=1$.  The size ratio 
 increases with $f$ when the excluded volume interactions between backbone and side poms are present.  
   The maximum impact on the size is observed in the case of SAW side poms, interacting with RW backbone ($\widetilde{w}_{ab}=3\epsilon/16$, $\widetilde{w}_a=\epsilon/8$).

An expression for  gyration radius of a single arm of rosette polymer structure reads:
\begin{eqnarray}
    &&\langle r^2_{g,\,{\rm arm}}\rangle_{r}=\frac{dL}{6}\left\{1+\frac{2\widetilde{u}_a}{\epsilon}-\frac{13}{12}\widetilde{u}_a+\frac{2}{5}\sqrt{5}f_cf_r\widetilde{w}_{ab}\left[\ln{2}-\ln\big(\sqrt{5}+3\big)+2\arctan\Big(1/\sqrt{5}\Big)\right]\right.\nonumber\\
    &&\left.-\frac{1}{25}f_r\widetilde{w}_{ab}\left[27\arctan\Big(1/\sqrt{5}\Big)\sqrt{5}-40\right]-[(f_c-1)/8](48\ln{2}-35)\widetilde{w}_a\right\}.
    \label{rarm}
\end{eqnarray}
Again,  substituting it into (\ref{ratior}) leads to the following expression for the size ratio:
\begin{eqnarray}
    &&p^{r}_g=1+\frac{2}{5}\sqrt{5}f_cf_r\widetilde{w}_{ab}\left[\ln{2}-\ln\big(\sqrt{5}+3\big)+2\arctan\Big(1/\sqrt{5}\Big)\right]\nonumber\\
    &&-\frac{1}{25}f_r\widetilde{w}_{ab}\left[27 \arctan\Big(1/\sqrt{5}\Big)\right]\sqrt{5}-40-[(f_c-1)/8](48\ln{2}-35)\widetilde{w}_a.
\label{sizer}
\end{eqnarray}
Again, let us consider the case when the linear branches of rosette polymer are of a RW type ($\widetilde{u}_b=0$). The results of evaluation of the size ratio (\ref{sizer}) at different possible interactions within the remaining loop branches and between linear and loop branches are presented in figure~\ref{pg} (right-hand) as functions of the branching parameter $f$.  The maximum impact on the size ratio is observed in the case of SAW loop branches, interacting with RW linear branches.

\section{Numerical calculations}\label{num}

 We started with a lattice model of self-avoiding walks (SAW) and random walks (RW) on simple cubic lattice. To simulate a set of conformations we use a pivot algorithm \cite{Clisby10,Madras88}. Choosing a random knot on one of the trajectories we perform one of the symmetry operations to the part of the structure:

\begin{itemize}
    \item for a side arm of pom-pom and open trajectory in rosette, the operation is performed between the pivot point and the free end of chosen arm;
    \item for the backbone of the pom-pom, the operation is performed between the pivot point and end points of all the side arms of one of the poms effectively moving it as well;
    \item for the closed trajectory of the rosette, the operation is performed between the pivot point and a point on the trajectory for which the chosen operation the trajectory remains closed.
\end{itemize}

First, $20 N F$ operations are used to reach a starting conformation, then $10^6$ operations are performed for each trajectory length $N$ to receive observables by averaging them over the ensemble of the obtained conformations. We perform the calculations for the lengths up to  $N=150$ steps on each of the $F$ trajectories.

Scaling exponents are calculated by approximating the data with:
\begin{equation}
\ln (\langle R_x^2 \rangle) = 2\nu_x\ln N+A.
\end{equation}
Results for this approximations are provided in tables \ref{tab1} and \ref{tab2}. 
They  are in good agreement with previous results for diblock chain and ring copolymers \cite{vlahos1994,vlahos1995} as well as miktoarm stars \cite{Zifferer2010} and show that interactions between the trajectories do not influence the scaling exponents, although for a conclusive statement additional sets of simulations and a second order analytical calculations are needed. However, we can make a conclusion that even in the higher order calculations $\widetilde{w}_{ab}$ interactions do not play any significant role, if any at all, in the scaling behaviour of copolymers.

\begin{table}[]
    \caption{Scaling exponents for pom-pom polymers.}
	\label{tab1}
    \centering
    \begin{tabular}{|c|c|c|c|c|}
    \hline
    $f$ & $\nu_{eff}$ & $\nu_{eff}$ & $\nu_{ab}$ & $\nu_{ab}$\\
       &  $\widetilde{w}_{ab}\neq0$ & $\widetilde{w}_{ab}=0$&  $\widetilde{w}_{ab}\neq0$ & $\widetilde{w}_{ab}=0$\\\hline
    $1$&  $0.556(2)$  & $0.560(3)$	& $0.554(2)$&$0.559(3)$\\\hline
    $2$& $0.563(2)$  &  $0.566(4)$  &$0.557(3)$&$0.561(4)$\\\hline
    $3$& $0.559(2)$  & $0.572(2)$&	$0.550(4)$&$0.566(3)$	\\\hline
    $4$& $0.571(5)$  & $0.572(6)$&	$0.565(6)$&$0.563(7)$ \\\hline
    $5$& $0.563(3)$  & $0.565(3)$&	$0.556(4)$&$0.556(4)$\\\hline
    \end{tabular}
\end{table}

\begin{table}[]
	\caption{Scaling exponents for rosette polymers.}
	\label{tab2}
    \centering
    \begin{tabular}{|c|c|c|c|c|c|}
    \hline
    $f_c$ & $f_r$& $\nu_{eff}$ & $\nu_{eff}$ & $\nu_{ab}$ & $\nu_{ab}$\\
      & &  $\widetilde{w}_{ab}\neq0$ & $\widetilde{w}_{ab}=0$&  $\widetilde{w}_{ab}\neq0$ & $\widetilde{w}_{ab}=0$\\\hline
$1$&$1$&$0.538(3)$ &$0.545(3)$ &$0.533(2)$ &$0.544(4)$\\\hline
$2$&$1$&$0.518(7)$ &$0.528(6)$ &$0.528(3)$ &$0.537(6)$\\\hline
$3$&$1$&$0.509(8)$ &$0.513(4)$ &$0.525(3)$ &$0.527(4)$\\\hline
$4$&$1$&$0.505(9)$ &$0.504(3)$ &$0.533(4)$ &$0.526(4)$ \\\hline
$1$&$2$&$0.540(7)$ &$0.548(5)$ &$0.529(7)$ &$0.538(6)$\\\hline
$2$&$2$&$0.535(8)$ &$0.541(4)$ &$0.531(3)$ &$0.543(4)$\\\hline
$3$&$2$&$0.535(9)$ &$0.527(4)$ &$0.526(3)$ &$0.535(5)$\\\hline
$4$&$2$&$0.518(7)$ &$0.517(7)$ &$0.533(4)$ &$0.527(8)$\\\hline  
    \end{tabular}
\end{table}

\section{Conclusions}\label{con}

We analyzed  conformational properties of two representatives of a class of complex branched macromolecules: the { rosette} structure containing $f_c$ linear branches and $f_r$ closed loops grafted to the central core, and 
the symmetric { pom-pom} structure, consisting of a backbone linear chain terminated by two branching points with functionalities $f$.
We consider the constituent strands (branches) of these structures to be of two various species $a$ and $b$, so that one of the species can be in $\Theta$-state under a particular solvent and temperature condition, whereas the other one is in a good solvent regime.
 For a chain with the repulsive excluded volume interactions between monomers, the conformational properties  are perfectly captured by a model of self-avoiding random walk (SAW); whereas to analyze the behaviour of polymer chain in the regime of $\Theta$-solvent (Gaussian polymers), the model of random walk (RW) is exploited. 

Within the frames of analytical approach based on direct polymer renormalization scheme, we eva\-luated the set of parameters characterizing the size properties of constituent parts of complex topologies considered [radii of gyration of backbone (\ref{r_backbone}) and side poms (\ref{r_pom}) of pom-pom copolymers and radius of gyration of individual linear branch of rosette copolymer (\ref{rarm})]. To quantitatively estimate the impact of  interactions between constituent parts of macromolecules on their size characteristics, we introduced the size ratios (\ref{sizepp}) and (\ref{sizer}). Our results confirm the increase of size ratios in presence of excluded volume as compared with $\Theta$-state. The values of effective critical exponents [equations \ref{nup} and (\ref{nur})], governing the effective linear size measure behaviour of pom-pom and rosette structures, correspondingly, are evaluated as well, both analytically and numerically.

\section*{Appendix}
\renewcommand{\theequation}{A.\arabic{equation}}
\renewcommand{\thefigure}{A.\arabic{figure}}
\setcounter{figure}{0}

Here, we give examples of diagrammatic calculations. We consider a simple case of two chains with both restriction points on the same trajectory and the interaction points on different ones. There are three possible diagrams and their sum reads:
\begin{eqnarray}
&&\int^L_0\,\rd s\,\int^L_0\,\rd z\,\int^s_0\,\rd s_2\,\int^{s_2}_0\,\rd s_1\left[s_2-s_1-\frac{(s_2-s_1)^2}{s+z}\right](s+z)^{-\frac{d}{2}}\nonumber\\
&&+\int^L_0\,\rd s\,\int^L_0\,\rd z\,\int^L_s\,\rd s_2\,\int^{s}_0\,\rd s_1\left[s_2-s_1-\frac{(s-s_1)^2}{s+z}\right](s+z)^{-\frac{d}{2}}\nonumber\\
&&+\int^L_0\,\rd s\,\int^L_0\,\rd z\,\int^L_s\,\rd s_2\,\int^{s_2}_s\,\rd s_1\left(s_2-s_1\right)(s+z)^{-\frac{d}{2}}.
\end{eqnarray}
Since all three integrals contain the same factor $(s+z)^{-\frac{d}{2}}$ that does not depend on the restriction points, the expression can be rewritten as:
\begin{eqnarray}
&&\int^L_0\,\rd s\,\int^L_0\,\rd z\,\left[\int^s_0\,\rd s_2\,\int^{s_2}_0\,\rd s_1\left(s_2-s_1-\frac{(s_2-s_1)^2}{s+z}\right)\right.\nonumber\\
&&+\left.\int^L_s\,\rd s_2\,\int^{s}_0\,\rd s_1\left(s_2-s_1-\frac{(s-s_1)^2}{s+z}\right)+\int^L_s\,\rd s_2\,\int^{s_2}_s\,\rd s_1\left(s_2-s_1\right)\right](s+z)^{-\frac{d}{2}}.
\end{eqnarray}

\begin{figure}[t!]
\begin{center}
\includegraphics[width=53mm]{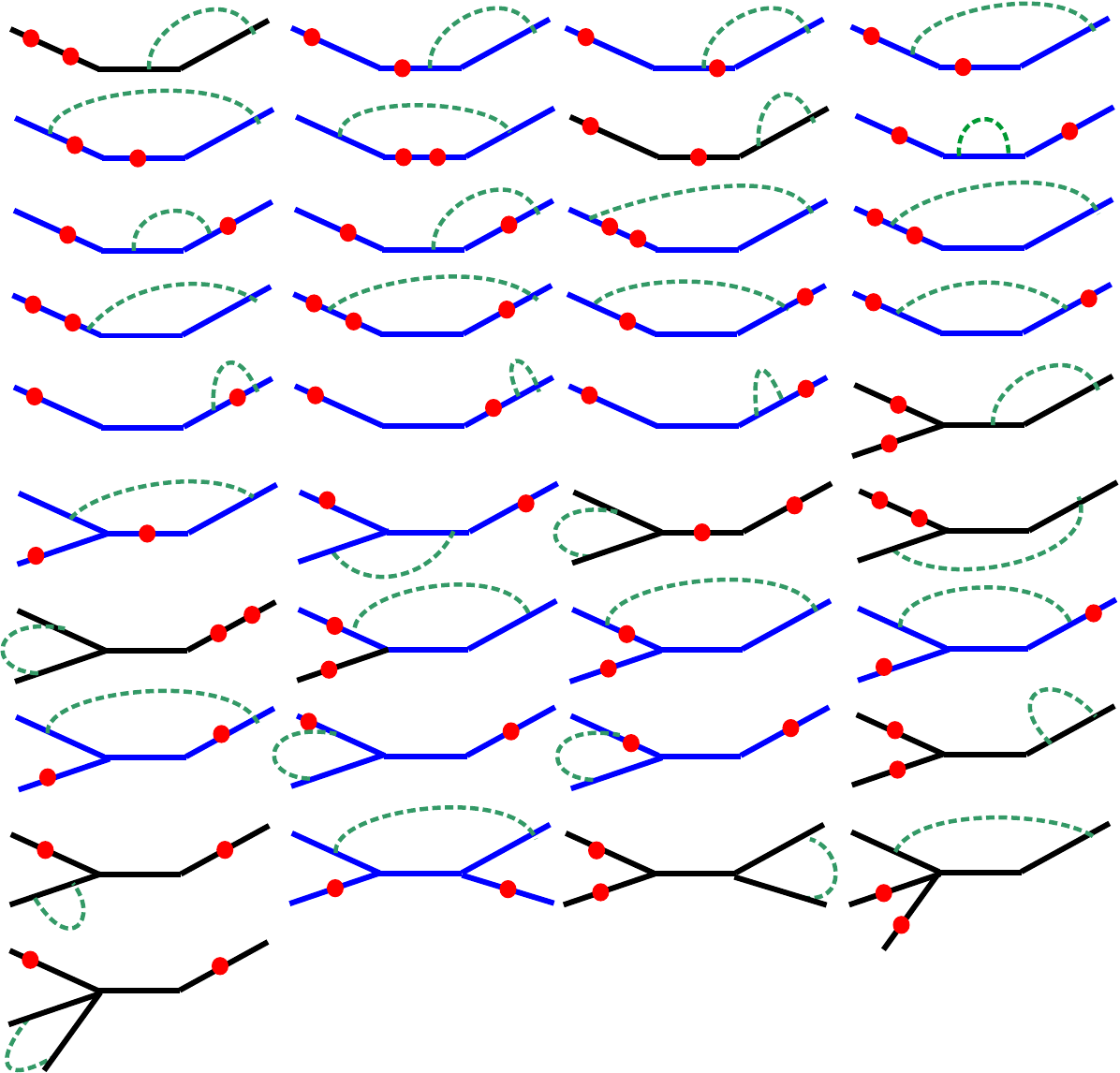}
\caption{(Colour online) Diagrammatic presentation of the contributions into the gyration radius in one loop approximation.}
 \label{Pom-pom_D_1l}
\end{center}
\end{figure}

\begin{figure}[b!]
\begin{center}
\includegraphics[width=45mm]{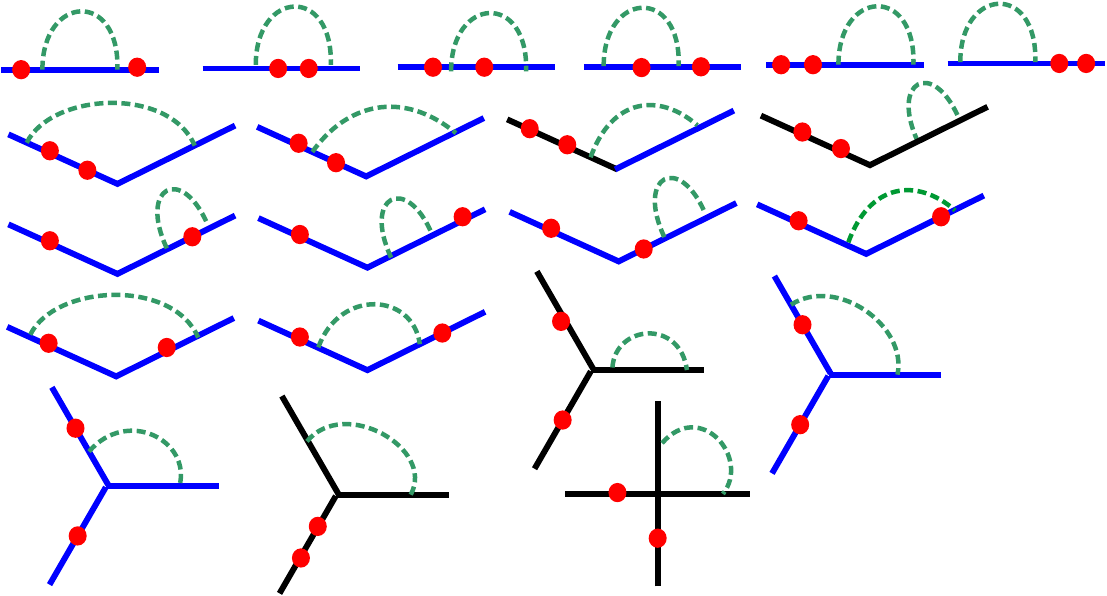}\includegraphics[width=45mm]{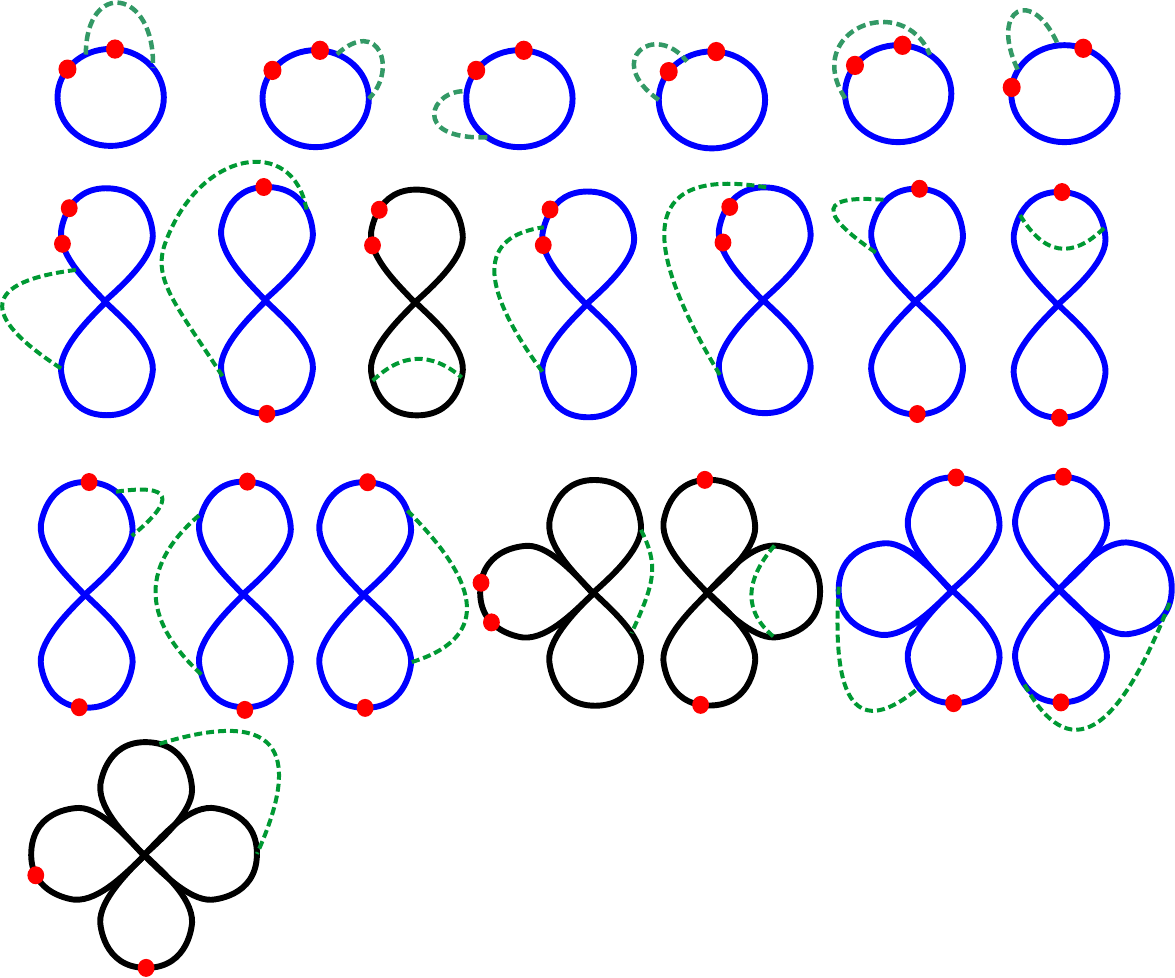}\includegraphics[width=45mm]{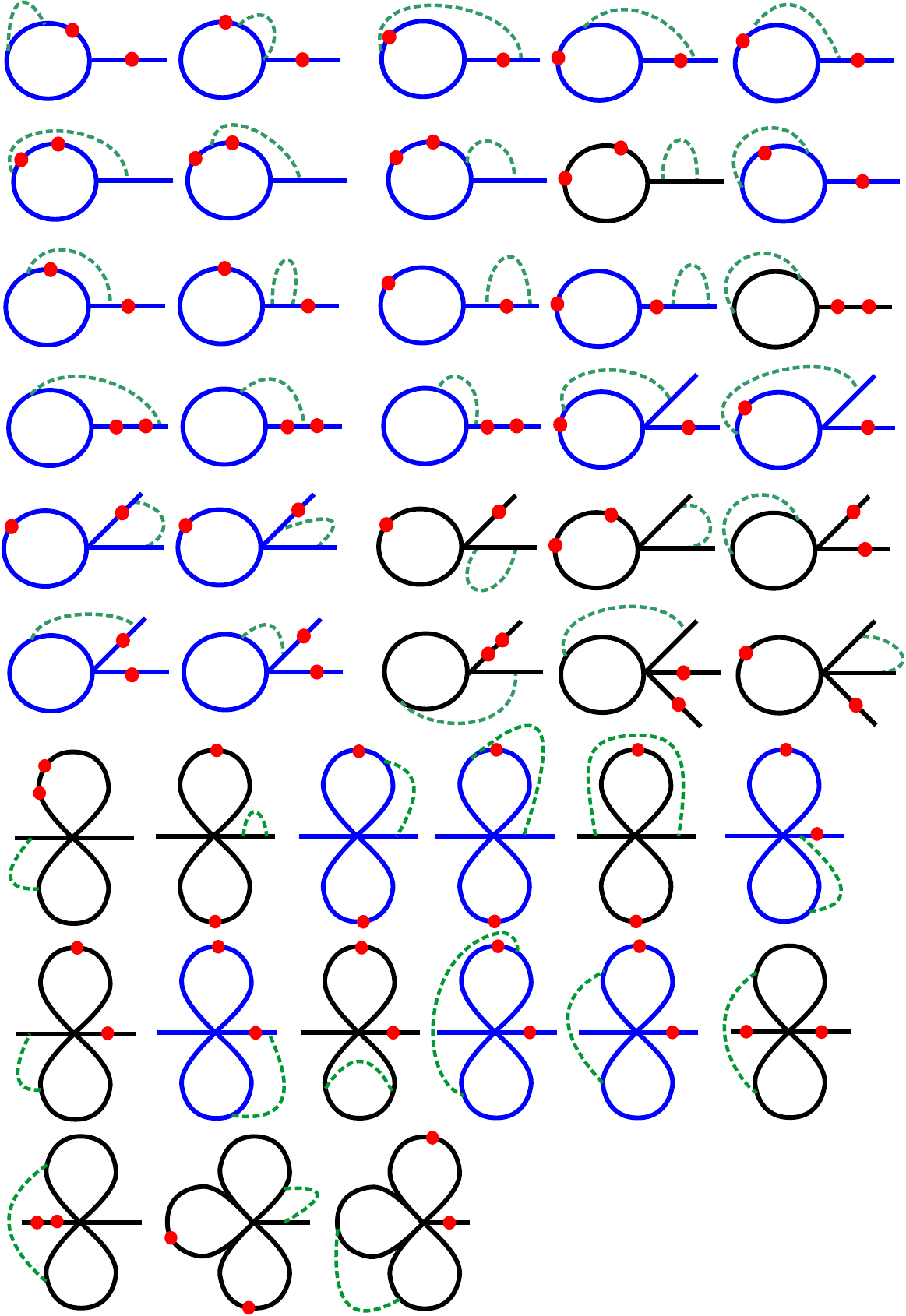}
\caption{(Colour online) Diagrammatic presentation of the contributions into the gyration radius in one loop approximation.}
 \label{Rossette_D_1l} 
\end{center}
\end{figure}

All the integrals inside $\left[ \ldots\right]$ contain the same term under the integration, which allows to rewrite the expression as:
\begin{eqnarray}
&&\int^L_0\,\rd s\,\int^L_0\,\rd z\,\left[\int^s_0\,\rd s_2\,\int^{s_2}_0\,\rd s_1\left(s_2-s_1\right)+\int^L_s\,\rd s_2\,\int^{s}_0\,\rd s_1\left(s_2-s_1\right)+\int^L_s\,\rd s_2\,\int^{s_2}_s\,\rd s_1\left(s_2-s_1\right)\right](s+z)^{-\frac{d}{2}}\nonumber\\
&&+\int^L_0\,\rd s\,\int^L_0\,\rd z\,\left[\int^s_0\,\rd s_2\,\int^{s_2}_0\,\rd s_1\left(-\frac{(s_2-s_1)^2}{s+z}\right)+\int^L_s\,\rd s_2\,\int^{s}_0\,\rd s_1\left(-\frac{(s-s_1)^2}{s+z}\right)\right](s+z)^{-\frac{d}{2}}.
\end{eqnarray}
The last two terms in the first line can be joined since the limits of the integration over $s_2$ are the same and the integration over $s_1$ can be presented as one integral:
\begin{eqnarray}
&&\int^L_0\,\rd s\,\int^L_0\,\rd z\,\left[\int^s_0\,\rd s_2\,\int^{s_2}_0\,\rd s_1\left(s_2-s_1\right)+\int^L_s\,\rd s_2\,\int^{s_2}_0\,\rd s_1\left(s_2-s_1\right)\right](s+z)^{-\frac{d}{2}}\nonumber\\
&&+\int^L_0\,\rd s\,\int^L_0\,\rd z\,\left[\int^s_0\,\rd s_2\,\int^{s_2}_0\,\rd s_1\left(-\frac{(s_2-s_1)^2}{s+z}\right)+\int^L_s\,\rd s_2\,\int^{s}_0\,\rd s_1\left(-\frac{(s-s_1)^2}{s+z}\right)\right](s+z)^{-\frac{d}{2}}.
\end{eqnarray}
The similar arguments now may be presented for the case of integration over $s_2$, so the final expression reads:
\begin{eqnarray}
&&\int^L_0\,\rd s\,\int^L_0\,\rd z\,(s+z)^{-\frac{d}{2}}\left[\int^L_0\,\rd s_2\,\int^{s_2}_0\,\rd s_1\left(s_2-s_1\right)\right]\nonumber\\
&&+\int^L_0\,\rd s\,\int^L_0\,\rd z\,\left[\int^s_0\,\rd s_2\,\int^{s_2}_0\,\rd s_1\left(-\frac{(s_2-s_1)^2}{s+z}\right)+\int^L_s\,\rd s_2\,\int^{s}_0\,\rd s_1\left(-\frac{(s-s_1)^2}{s+z}\right)\right](s+z)^{-\frac{d}{2}}.
\label{L7-9}
\end{eqnarray}
The first line here is a multiplication of the diagram $\xi_1$ and $Z_2$ from the pom-pom partition function. The calculation of the gyration radius in one loop approximation in general may be presented as:
\begin{eqnarray}
&&\langle R^2_g\rangle = Z^{-1}(\langle R^2_g\rangle_0 - u (\textrm{Sum of digrams}))=\langle R^2_g\rangle_0Z^{-1}\left(1-u\frac{(\textrm{Sum of digrams})}{\langle R^2_g\rangle_0}\right)\nonumber\\
&&=\langle R^2_g\rangle_0(1+uZ_x)\left(1-u\frac{(\textrm{Sum of digrams})}{\langle R^2_g\rangle_0}\right)=\langle R^2_g\rangle_0\left(1-u\left[\frac{(\textrm{Sum of digrams})}{\langle R^2_g\rangle_0}-Z_x\right]\right).
\end{eqnarray}
The expression in square brackets can be rewritten as $\displaystyle\frac{(\textrm{Sum of digrams})-Z_x\langle R^2_g\rangle_0}{\langle R^2_g\rangle_0}$. Note that all the diagrams can be divided into two groups: the reducible one presented in figures~\ref{Pom-pom_D_1l} and \ref{Rossette_D_1l} with the black lines [their contributions are always presented as the product of the diagrams {contributions denoted by $\xi$ (shown in figure~\ref{fig:3}) and denoted by $Z$ (shown in figures~\ref{fig:2} and \ref{fig:2a})}] and irreducible are  presented with the blue lines. The latter group can be divided into smaller groups for each of which the arguments conducted above can be provided, thus allowing to separate the contributions of the type $\xi_yZ_x$, {where $\xi_y$ is a contribution into the gyration radius in Gaussian approximation and $Z_x$ }on a much earlier stage. These contributions are cancelled by $Z_x\langle R^2_g\rangle_0$ and thus we do not need to calculate them. We can limit the calculations to considering the integrals that do not get cancelled, which considerably reduces the calculations. Since the groups of diagrams that contain the interactions between different trajectories do not contribute into the scaling exponents or, in other words, do not contain poles in the $\epsilon$-expansions, and if necessary can be calculated at a fixed space dimension $d=3$ when considering Douglas--Freed approximation, only these terms will be important in one loop approximation.

%\newpage
\ukrainianpart

\title{Універсальні властивості галужених кополімерів у слабких розчинах }

\author[Х. Гайдуківська, В. Блавацька ]
{Х. Гайдуківська \refaddr{label1,label2},
	В. Блавацька \refaddr{label1,label3}
}

\addresses{
	\addr{label1} Iнститут фiзики конденсованих систем Нацiональної академiї наук України, вул. Свєнцiцького 1, 79011 Львiв, Україна
	\addr{label2}  Iнститут Фiзики, Сiлезький унiверситет, вул. Першого полку пiхоти 75, 41-500 Хоржув, Польща
	\addr{label3} Дiоскурi центр фiзики i хiмiї бактерiй, Iнститут фiзичної хiмiї, Польська академiя наук, 01-224 Варшава, Польща
}

\makeukrtitle
\begin{abstract}
	\tolerance=3000%
	
	Проаналізовано універсальні конформаційні властивості складних полімерних макромолекул на основі двох топологій: структура {\it розетки}, що містить $f_c$ лінійних гілок і $f_r$ замкнених петель, прикріплених до центрального кору, і структура симетричного {\it пом-пом}, що складається із основного лінійного ланцюжка із двома точками галуження функціональності $f$ на обох кінцях. Вважається, що складові ланки (гілки) цих структур можуть бути двох різних типів хімічного складу    $a$ та $b$.  Залежно від умов розчинника,  взаємодії в мехаж ланцюжків та між ланцюжками можуть зникати, що відповідає  $\Theta$-стану відповідного типу полімера.  Застосовуючи як аналітичний підхід в рамках прямого полімерного перенормування, так і чисельні симуляції на основі граткової моделі полімера, отримано набір параметрів, що характеризують розмірні характеристики складових частин обох складних топологій, і кількісно описано вплив взаємодій між складовими частинами на ці характеристики.
\end{abstract}

\end{document}